\documentclass[twocolumn]{aastex6}
\usepackage{kotexutf}

\def\adfs27{ADFS-27}
\def\adfsn{ADFS-27N}
\def\adfss{ADFS-27S}

\received{05/25/2017}
\revised{08/31/2017}
\accepted{09/12/2017}

\usepackage{apjfonts}
\usepackage{aas_macros, xspace} 
\newcommand{\eq}{\,=\,}

\def\ts     {\thinspace}
\def\kms    {\ts km\ts s$^{-1}$}

\def\msol   {$M_{\odot}$}
\def\lsol   {$L_{\odot}$}
\def\lprime {K\,\kms\,pc$^2$}

\def\aco    {{\rm CO}($J$=1$\to$0)}

\def\eco    {{\rm CO}($J$=5$\to$4)}
\def\fco    {{\rm CO}($J$=6$\to$5)}

\def\water   {{\rm H$_2$O}(2$_{11}$$\to$2$_{02}$)}

\shorttitle{An Extremely Red Binary HyLIRG at $z$$\sim$6}
\shortauthors{Riechers et al.}

\begin{document}

\title{Rise of the Titans:\ A Dusty, Hyper-Luminous ``870\,$\mu$m Riser'' Galaxy at $z$$\sim$6}


\author{Dominik A.\ Riechers\altaffilmark{1}}
\author{T.~K.\ Daisy Leung\altaffilmark{1}}
\author{Rob J.\ Ivison\altaffilmark{2,3}}
\author{Ismael P\'erez-Fournon\altaffilmark{4,5}}
\author{Alexander J.~R.\ Lewis\altaffilmark{3}}
\author{Rui~Marques-Chaves\altaffilmark{4,5}}
\author{Iv\'an Oteo\altaffilmark{2,3}}
\author{Dave L.\ Clements\altaffilmark{6}}
\author{Asantha Cooray\altaffilmark{7}}
\author{Josh Greenslade\altaffilmark{6}}
\author{Paloma~Mart\'inez-Navajas\altaffilmark{4,5}}
\author{Seb Oliver\altaffilmark{8}}
\author{Dimitra Rigopoulou\altaffilmark{9,10}}
\author{Douglas Scott\altaffilmark{11}}    
\author{Axel Wei\ss\altaffilmark{12}}

\altaffiltext{1}{Cornell University, Space
 Sciences Building, Ithaca, NY 14853, USA}
\altaffiltext{2}{European Southern Observatory,
  Karl-Schwarzschild-Stra{\ss}e 2, D-85748 Garching, Germany}
\altaffiltext{3}{Institute for Astronomy, University of Edinburgh,
  Royal Observatory, Blackford Hill, Edinburgh EH9 3HJ, UK}
\altaffiltext{4}{Instituto de Astrofisica de Canarias, E-38200 La
  Laguna, Tenerife, Spain}
\altaffiltext{5}{Departamento de Astrofisica, Universidad de La
  Laguna, E-38205 La Laguna, Tenerife, Spain}
\altaffiltext{6}{Astrophysics Group, Imperial College London, Blackett
  Laboratory, Prince Consort Road, London SW7 2AZ, UK}
\altaffiltext{7}{Department of Physics and Astronomy, University of
  California, Irvine, CA 92697, USA}
\altaffiltext{8}{Astronomy Centre, Department of Physics and
  Astronomy, University of Sussex, Brighton BN1 9QH, UK}
\altaffiltext{9}{Department of Physics, University of Oxford, Keble
  Road, Oxford OX1 3RH, UK}
\altaffiltext{10}{Space Science and Technology Department, Rutherford
  Appleton Laboratory, Chilton, Didcot, Oxfordshire OX11 0QX, UK}
\altaffiltext{11}{Department of Physics and Astronomy, University of
  British Columbia, 6224 Agricultural Road, Vancouver, BC V6T 1Z1,
  Canada}
\altaffiltext{12}{Max-Planck-Institut f\"ur Radioastronomie, Auf dem
  H\"ugel 69, D-53121 Bonn, Germany}

 \email{riechers@cornell.edu}

\begin{abstract}

We report the detection of \adfs27, a dusty, starbursting major merger
at a redshift of $z$=5.655, using the Atacama Large
Millimeter/submillimeter Array (ALMA). \adfs27\ was selected from {\em
  Herschel}/SPIRE and APEX/LABOCA data as an extremely red
``870\,$\mu$m riser'' (i.e., $S_{250\mu{\rm m}}$$<$$S_{350\mu{\rm
    m}}$$<$$S_{500\mu{\rm m}}$$<$$S_{870\mu{\rm m}}$), demonstrating
the utility of this technique to identify some of the highest-redshift
dusty galaxies. A scan of the 3\,mm atmospheric window with ALMA
yields detections of \eco\ and \fco\ emission, and a tentative
detection of \water\ emission, which provides an unambiguous redshift
measurement. The strength of the CO lines implies a large molecular
gas reservoir with a mass of $M_{\rm
  gas}$=2.5$\times$10$^{11}$\,$(\alpha_{\rm
  CO}/0.8)\,(0.39/r_{51})$\,\msol, sufficient to maintain its
$\sim$2400\,\msol\,yr$^{-1}$ starburst for at least
$\sim$100\,Myr. The 870\,$\mu$m dust continuum emission is resolved
into two components, 1.8 and 2.1\,kpc in diameter, separated by
9.0\,kpc, with comparable dust luminosities, suggesting an ongoing
major merger. The infrared luminosity of $L_{\rm
  IR}$$\simeq$2.4$\times$10$^{13}$\,\lsol\ implies that this system
represents a binary hyper-luminous infrared galaxy, the most distant
of its kind presently known. This also implies star formation rate
surface densities of $\Sigma_{\rm SFR}$=730 and
750\,\msol\,yr$^{-1}$\,kpc$^2$, consistent with a binary ``maximum
starburst''. The discovery of this rare system is consistent with a
significantly higher space density than previously thought for the
most luminous dusty starbursts within the first billion years of
cosmic time, easing tensions regarding the space densities of
$z$$\sim$6 quasars and massive quiescent galaxies at $z$$\gtrsim$3.

\end{abstract}


\section{Introduction} \label{sec:intro}

Detailed studies of dusty star-forming galaxies (DSFGs) at high
redshift selected at (sub-)millimeter wavelengths (submillimeter
galaxies, or SMGs) over the past two decades have shown them to be a
key ingredient in our understanding of the early formation of massive
galaxies (see \citealt{blain02a,casey14a} for reviews). The brightest,
``hyper-luminous'' DSFGs (hyper-luminous infrared galaxies, or
HyLIRGs) represent some of the most luminous, massive galaxies in the
early universe, reaching infrared luminosities of $L_{\rm
  IR}$$>$10$^{13}$\,\lsol, and star formation rates in excess of
1000\,\msol\,yr$^{-1}$, emerging from compact regions only few
kiloparsec in diameter
\citep[e.g.,][]{riechers13b,riechers14b,fu13,ivison13,hodge15,hodge16,oteo16}.
While the general DSFG population is thought to be somewhat
heterogeneous
\citep[e.g.,][]{dave10,narayanan10b,narayanan15,hayward12}, these
HyLIRGs are likely major mergers of gas-rich galaxies
\citep[e.g.,][]{engel10,riechers11e,ivison11,ivison13,oteo16}, and
they may also be associated with protoclusters of galaxies, which
represent some of the most overdense environments in the early
universe \citep[e.g.,][]{daddi09,capak11,ivison13}.

Due to their high dust content, it is common that most of the stellar
light in DSFGs is subject to dust extinction, rendering their
identification out to the highest redshifts notoriously
difficult. While many DSFGs were found at $z$=2--3.5 relatively early
on (e.g., \citealt{ivison98,ivison00,chapman05}), more than a decade
passed between the initial discovery of this galaxy population and the
identification of the first examples at $z$$>$4
\citep{capak08,daddi09} and $z$$>$5
\citep{riechers10a,capak11,walter12}.

Once the {\em Herschel Space Observatory} was launched, it became
possible to develop color selection techniques to systematically
search for the most distant DSFGs in large-area surveys like the
Herschel Multi-tiered Extragalactic Survey \citep[HerMES;
][]{oliver12}.  Since the peak of the far-infrared (FIR) spectral
energy distribution (SED) shifts through the 250, 350 and 500\,$\mu$m
bands probed by {\em Herschel's} Spectral and Photometric Imaging
Receiver (SPIRE), the most distant sources typically appear ``red''
between these bands, i.e., $S_{250\mu{\rm m}}$$<$$S_{350\mu{\rm
    m}}$$<$$S_{500\mu{\rm m}}$, with steeper (``ultra-red'') color
criteria resulting in the selection of potentially more distant
sources \citep[e.g.,][]{riechers13b,ivison16}. Based on FIR
photometric redshift estimates, the median redshifts of these sources
have been suggested to be $\langle$$z$$\rangle$\eq3.7 to 4.7, where
different redshift values are obtained for different samples due to
the exact color cutoffs, flux density limits, and redshift fitting
techniques chosen (e.g., \citealt{dowell14,ivison16}; see also
\citealt{asboth16}). Spectroscopic confirmation of a subsample of 25
sources based on CO rotational lines, an indicator of the molecular
gas that fuels the intense star formation in these systems (see
\citealt{cw13} for a review), has verified the higher median redshifts
compared to general DSFG samples (e.g.,
\citealt{cox11,combes12,riechers13b}; D.~Riechers et al., in prep.;
Fudamoto et al., in prep.). These studies find redshifts as high as
$z$\eq6.34 (\citealt{riechers13b}).
In an alternative approach, surveys with the South Pole Telescope
(SPT) have revealed a sample of gravitationally-lensed DSFGs selected
at 1.4 and 2\,mm with a spectroscopic median redshift of
$\langle$$z$$\rangle$\eq3.9 \citep[e.g.,][]{weiss13,strandet16}. A
substantial fraction of this sample would also fulfill {\em
  Herschel}-red sample selection criteria.

With this paper, we aim to extend the {\em Herschel}-red and ultra-red
criteria through the identification of ``extremely red'' DSFGs with
$S_{250\mu{\rm m}}$$<$$S_{350\mu{\rm m}}$$<$$S_{500\mu{\rm
    m}}$$<$$S_{870\mu{\rm m}}$. Such ``870\,$\mu$m riser'' galaxies
should, in principle, lie at even higher redshifts than the bulk of
the red DSFG population. We here present detailed follow-up
observations of the first such source we have identified in the {\em
  Herschel} HerMES data, 2HERMES S250 SF J043657.7--543810
(hereafter:\ \adfs27).  We use a concordance, flat $\Lambda$CDM
cosmology throughout, with $H_0$\eq69.6\,\kms\,Mpc$^{-1}$,
$\Omega_{\rm M}$\eq0.286, and $\Omega_{\Lambda}$\eq0.714.

\section{Data} \label{sec:data}

\subsection{Herschel/PACS+SPIRE}

\adfs27\ was observed with the {\em Herschel Space Observatory} as
part of HerMES, covering 7.47\,deg$^2$ in the Akari Deep Field South
(ADFS). The field was observed for 18.1\,hr with the PACS and SPIRE
instruments in parallel mode, resulting in nominal instrumental noise
levels of 49.9, 95.1, 25.8, 21.2, and 30.8\,mJy (5$\sigma$ rms) at
110, 160, 250, 350, and 500\,$\mu$m, respectively.\footnote{Quoted
  sensitivities are single-pixel rms values, which are worse than the
  flux uncertainties of point sources achieved after employing matched
  filtering techniques (e.g., \citealt{oliver12,schulz17}).} The flux
scale is accurate to $\sim$5\%. \adfs27 was detected at 250, 350, and
500\,$\mu$m, but not shortwards. Flux densities were extracted using
{\em Starfinder} and {\em SussExtractor}, and from the band-merged
xID250 catalog published as part of HerMES DR4. This yields $S_{\rm
  250\mu{\rm m}}$=(14.3$\pm$2.3), (13.0$\pm$2.6), and
(14.3$\pm$2.3)\,mJy, $S_{\rm 350\mu{\rm m}}$\eq(20.3$\pm$2.4),
(18.5$\pm$2.5), and (19.1$\pm$2.3)\,mJy, and $S_{\rm 500\mu{\rm
    m}}$\eq(22.0$\pm$2.6), (22.2$\pm$2.9), and (24.0$\pm$2.7)\,mJy,
respectively. These uncertainties do not include the contribution due
to source confusion, which typically dominates. We however note that
the source is relatively isolated in the SPIRE maps
(Fig.~\ref{f1}). xID250-based flux densities are adopted in the
following (Table~\ref{t1}).  From these data, \adfs27\ was selected as
a ``red source'' (i.e., $S_{250\mu{\rm m}}$$<$$S_{350\mu{\rm
    m}}$$<$$S_{500\mu{\rm m}}$) for further follow-up observations.

\subsection{APEX/LABOCA}

\begin{figure*}
\epsscale{1.15}
\plotone{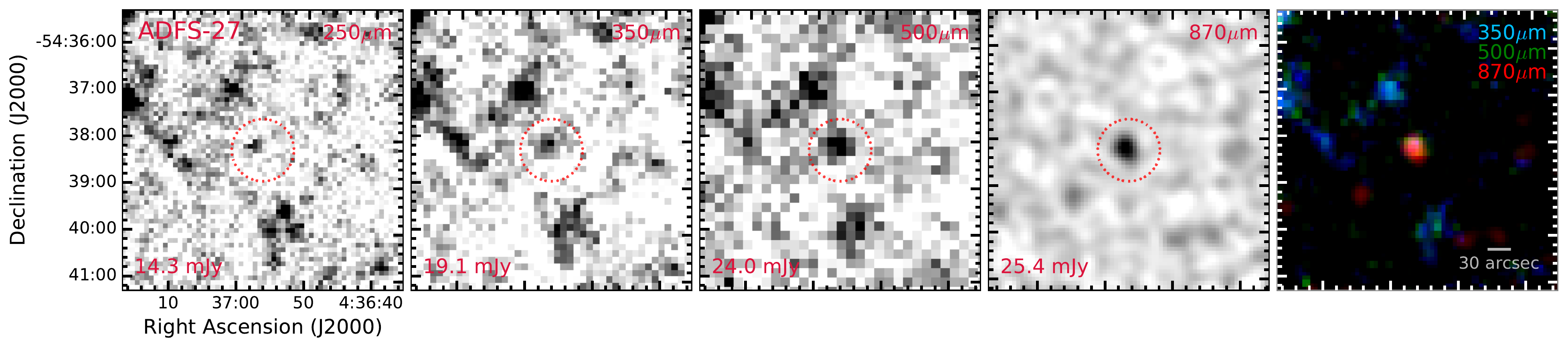}
\vspace{-2mm}

\caption{{\em Herschel}/SPIRE 250, 350, and 500\,$\mu$m and
  APEX/LABOCA 870\,$\mu$m images centered on \adfs27, and
  870/500/350\,$\mu$m color composite ({\em left} to {\em
    right}). Source flux densities are indicated in the bottom left
  corners of the first four panels (see Table~\ref{f1} for
  uncertainties). The source is relatively isolated in the deep SPIRE
  maps.\label{f1}}
%
\end{figure*}

We observed \adfs27\ at 870\,$\mu$m with the Large APEX bolometer
camera (LABOCA) mounted on the 12 m Atacama Pathfinder EXperiment
(APEX) telescope. Observations were carried out on 2012 September 17
as part of program M-090.F-0025-2012, resulting in 3.4\,hr on source
time. Individual scans had a length of $\sim$7\,min, resulting in a
map that fully samples the $\sim$11\,arcmin diameter field-of-view of
LABOCA. Pointing was checked on nearby quasars every hour, and was
stable to within $\sim$3$''$ rms. The effective FWHM beam size, as
measured on the pointing source J2258--280, was 19.2$''$. Precipitable
water vapor columns varied between 0.4 and 1.3\,mm, corresponding to
zenith atmospheric opacities of 0.2--0.4 in the LABOCA passband. This
resulted in an rms noise level of 1.8\,mJy\,beam$^{-1}$ at the
position of \adfs27\ (3.7\,mJy\,beam$^{-1}$ map average) in a map
smoothed to 27$''$ resolution. The flux density scale was determined
through observations of Uranus and Neptune, yielding an accuracy of
$\sim$7\%. Data reduction was performed with the {\em BoA} package,
applying standard calibration techniques. These observations were used
to select \adfs27\ as an ``extremely red'' source with $S_{250\mu{\rm
    m}}$$<$$S_{350\mu{\rm m}}$$<$$S_{500\mu{\rm m}}$$<$$S_{870\mu{\rm
    m}}$ (Fig.~\ref{f1}; Table~\ref{t1}).

\subsection{ALMA 870\,$\mu$m}

We observed 870\,$\mu$m continuum emission toward \adfs27\ using ALMA
(project ID:\ 2013.1.00001.S; PI:\ Ivison). Observations were carried
out on 2015 August 31 with 33 usable 12\,m antennas under good weather
conditions in an extended array configuration (baseline
range:\ 15--1466\,m). This resulted in 5.1\,min of usable on source
time, centered on the {\em Herschel}/SPIRE 500\,$\mu$m position. The
nearby quasar J0425--5331 was observed regularly for pointing,
amplitude and phase calibration, while J0538--4405 was observed for
bandpass calibration, and J0519--4546 was used for absolute flux
calibration, leading to $<$10\% calibration uncertainty. The
correlator was set up with two spectral windows of 1.875\,GHz
bandwidth (dual polarization) each per sideband, centered at a local
oscillator frequency of 343.463325\,GHz, with a frequency gap of
8\,GHz between the sidebands.

Data reduction was performed using version 4.7.1 of the Common
Astronomy Software Applications ({\sc casa}) package. Data were mapped
using the CLEAN algorithm with ``natural'' and robust 0.5 weighting,
resulting in synthesized beam sizes of 0\farcs20$\times$0\farcs17 and
0\farcs17$\times$0\farcs14 at rms noise values of 99 and
108\,$\mu$Jy\,beam$^{-1}$ in the phase center over the entire 7.5\,GHz
bandwidth, respectively. Due to its distance from the phase center,
the noise is increased by a primary beam attenuation factor of 1.62 at
the position of \adfs27.

\subsection{ALMA 3\,mm}

We scanned the
84.077033--113.280277\,GHz frequency range to search for spectral
lines toward \adfs27\ using ALMA (project ID:\ 2016.1.00613.S;
PI:\ Riechers). Observations were carried out under good weather
conditions during six runs between 2017 January 5 and 9 with 40--47
usable 12\,m antennas in a compact array configuration (baseline
range:\ 15--460\,m). We used five spectral setups, resulting in a
total on source time of 45.7\,min (7.8--14.1\,min per setup), centered
on the ALMA 870\,$\mu$m position. The nearby quasar J0425--5331 was
observed regularly for pointing, amplitude and phase
calibration. J0519--4546 was used for bandpass and absolute flux
calibration, leading to $<$10\% calibration uncertainty.

The correlator was set up with two spectral windows of 1.875\,GHz
bandwidth (dual polarization) each per sideband, at a sideband
separation of 8\,GHz. Full frequency coverage was attained by shifting
setups in frequency by $\sim$3.75\,GHz, such that subsequent settings
filled in part of the IF gap in the first spectral setup. This allowed
us to cover the full range of $\sim$29.21\,GHz without significant
gaps in frequency, but resulted in some frequency overlap near
97.5\,GHz (see Fig.~\ref{f2} for effective exposure times across the
full band).

Data reduction was performed using version 4.7.1 of the {\sc casa}
package. Data were mapped using the CLEAN algorithm with ``natural''
and robust 0.5 weighting, resulting in synthesized beam sizes of
3\farcs13$\times$2\farcs36 and 2\farcs48$\times$1\farcs86 at rms noise
values of 11.2 and 13.6\,$\mu$Jy\,beam$^{-1}$ in the phase center over
a line-free bandwidth of 27.40\,GHz after averaging all spectral
setups, respectively. Spectral line cubes mapped with ``natural''
weighting at 86.6, 103.9, and 113.0\,GHz yield beam sizes of
3\farcs68$\times$2\farcs72, 3\farcs05$\times$2\farcs26, and
2\farcs83$\times$2\farcs17 at rms noise levels of 352, 509, and
297\,$\mu$Jy\,beam$^{-1}$ per 19.55, 19.55, and 58.65\,MHz bin,
respectively. Imaging the same data at 103.9\,GHz with robust --2
(``uniform'') weighting yields a beam size of
2\farcs11$\times$1\farcs56 at $\sim$1.9 times higher rms noise.

\subsection{Spitzer/IRAC}

\adfs27\ was covered with {\em Spitzer}/IRAC at 3.6 and 4.5\,$\mu$m
between 2011 November 17--21 (program ID:\ 80039; PI:\ Scarlata) and
targeted for deeper observations on 2015 May 24 (program ID:\ 11107;
PI:\ Perez-Fournon). Data reduction was performed using the {\em
  MOPEX} package using standard procedures. Absolute astrometry was
obtained relative to Gaia DR1, yielding rms accuracies of 0.04$''$ and
0.06$''$ in the 3.6 and 4.5\,$\mu$m bands, respectively. Photometry
was obtained with the {\em SExtractor} package, after de-blending from
two foreground objects and sky removal using {\em GALFIT.}

\subsection{VISTA and WISE}

The position of \adfs27\ was covered by the VISTA Hemisphere Survey
(VHS) DR4 on 2010 November 19 and by the {\em Wide-field Infrared
  Survey Explorer} ({\em WISE}) as part of the allWISE survey between
2010 January 19 and 2011 January 30. \adfs27\ is not detected in the
VHS 1.25, 1.65, and 2.15\,$\mu$m ($J$/$H$/$K_{\rm s}$) bands. It is
strongly blended with a nearby star ($m_{\rm gaia}$=18.20) in the 3.4
and 4.6\,$\mu$m (W1 and W2) bands, such that no useful limit can be
obtained. It also remains undetected in the 12 and 22\,$\mu$m (W3 and
W4) bands.

\begin{figure*}
\epsscale{1.15}
\plotone{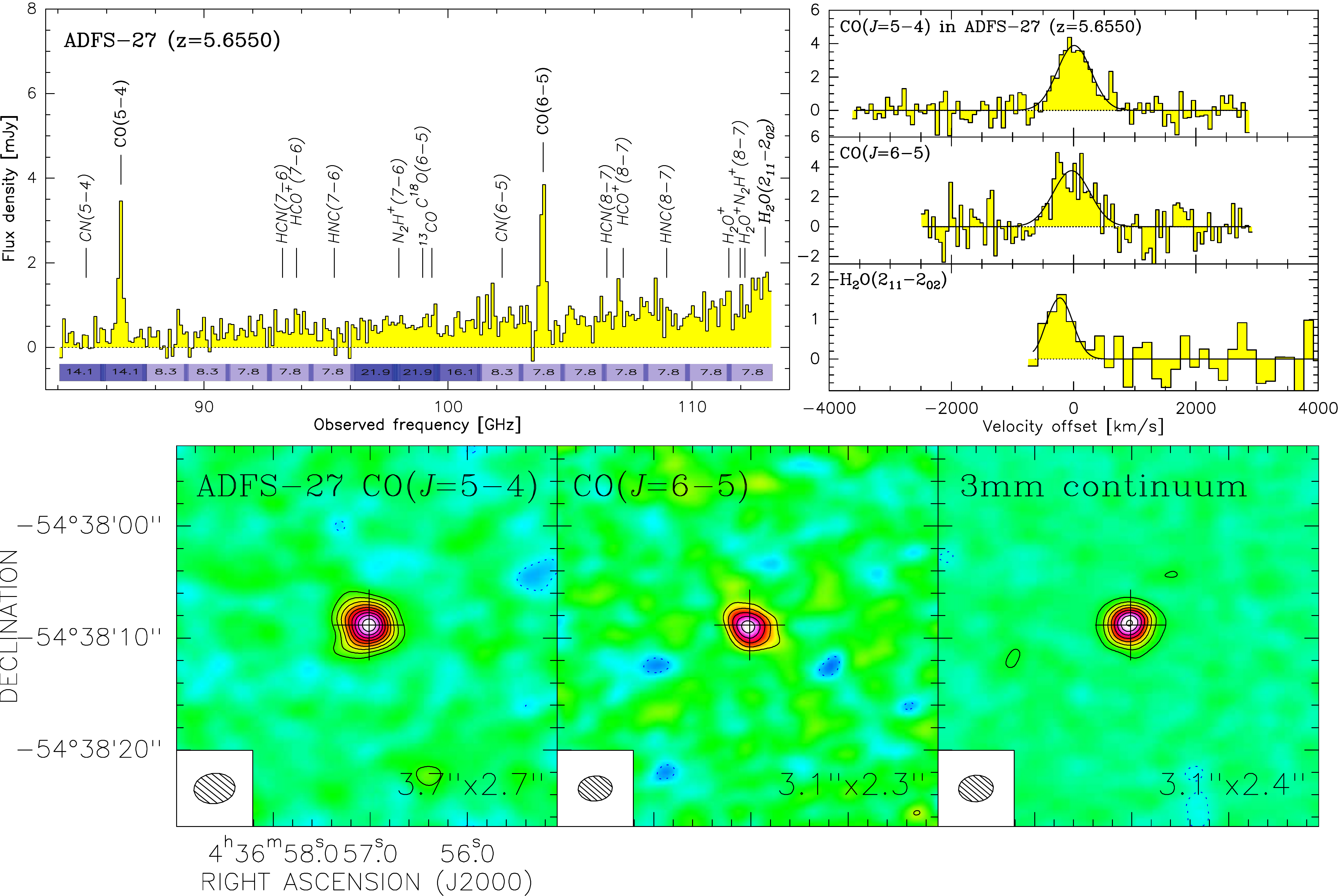}
\vspace{-2mm}

\caption{ALMA 3\,mm spectra ({\em top}) and maps ({\em bottom}) of the
  line and continuum emission toward \adfs27. {\em Top left:} Full
  spectrum obtained after combination of all spectral setups at a
  spectral resolution of 117.3\,MHz. The stripe near the bottom shows
  the integration time in minutes in each 1.875\,GHz spectral
  window. Increasingly darker colors indicate regions covered by two
  or three observing runs due to repetition or tuning overlap. {\em
    Top Right:} Zoom-in regions showing the spectral lines used for
  the redshift identification after continuum subtraction. The
  \water\ line is only marginally detected. Spectra of the \eco, \fco,
  and \water\ lines are shown at spectral resolutions of 19.55, 19.55,
  and 58.65\,MHz (68, 56, and 156\,\kms ), respectively. {\em Bottom:}
  Integrated line maps of the \eco\ and \fco\ emission over 651 and
  711\,\kms, and continuum map across the line-free spectral range
  (27.40\,GHz), imaged with natural baseline weighting. The beam sizes
  are indicated in the bottom left corner of each panel. Contours are
  shown in steps of 2$\sigma$ (lines) and 5$\sigma$ (continuum),
  starting at 3$\sigma$, where 1$\sigma$=0.084\,Jy\,\kms\,beam$^{-1}$,
  0.12\,Jy\,\kms\,beam$^{-1}$, and 11.2\,$\mu$Jy\,beam$^{-1}$,
  respectively. The cross in each panel indicates the peak position of
  the \eco\ emission. \label{f2}}
%
\end{figure*}

\section{Results}

\subsection{Continuum Emission}

\begin{figure*}
\epsscale{1.15}
\plotone{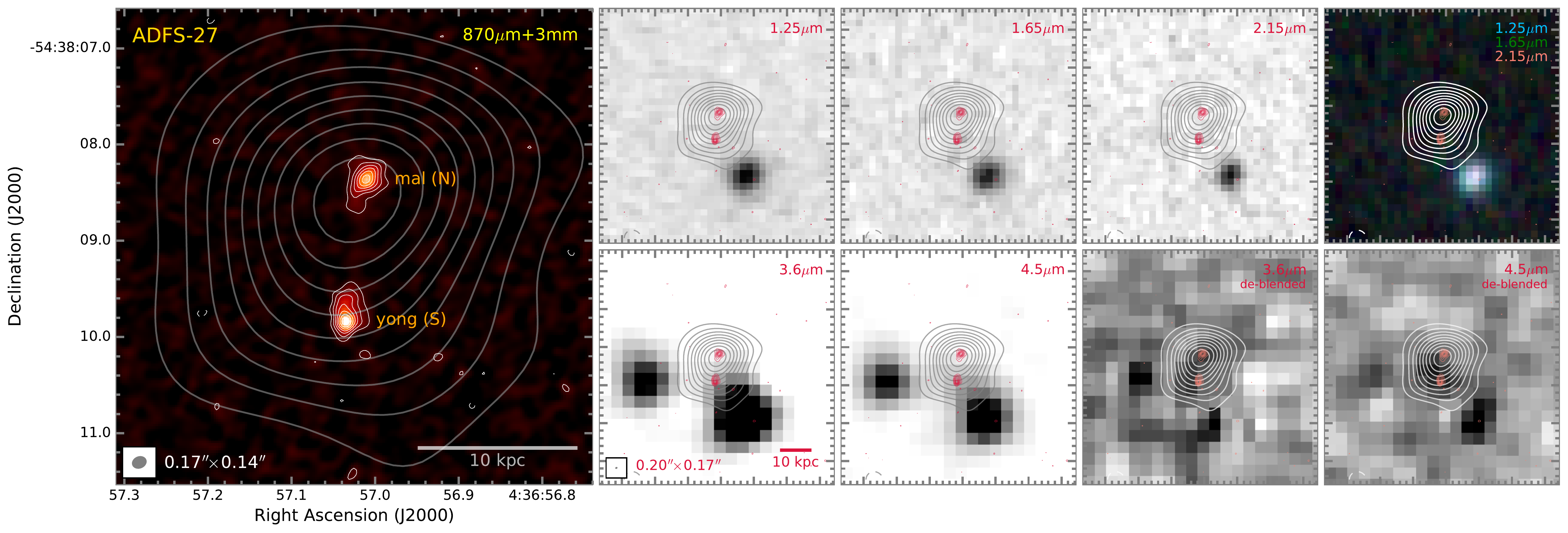}
\vspace{-2mm}

\caption{ALMA 870\,$\mu$m, {\em Spitzer}/IRAC 3.6 and 4.5\,$\mu$m, and
  VISTA 1.25, 1.65, and 2.15\,$\mu$m imaging of \adfs27. {\em Left:}
  870\,$\mu$m imaging (color scale and white contours) overlaid with
  3\,mm continuum contours (gray). Both data sets are imaged with
  robust 0.5 weighting. {\em Bottom middle:} 3.6 and 4.5\,$\mu$m
  imaging overlaid with 870\,$\mu$m (crimson color, natural weighting)
  and 3\,mm continuum contours (same as left panel). {\em Bottom
    right:} Same as middle, but with foreground sources subtracted and
  contrast adjusted. {\em Top middle/right:} VISTA images and color
  composite with the same contours as bottom. The bright blue source
  to the south-west of \adfs27\ is a star detected by {\em
    Gaia}. 870\,$\mu$m beam sizes are indicated in the bottom left
  corners of two panels. 3\,mm beam size is
  2\farcs48$\times$1\farcs86. Contours are in steps of $\pm$3$\sigma$,
  where 1$\sigma$=99, 108, and 13.6\,$\mu$Jy\,beam$^{-1}$ in the phase
  center for the robust 0.5 and natural weighting 870\,$\mu$m and
  3\,mm data, respectively.
  \label{f3}}
%
\end{figure*}

\begin{figure*}
\epsscale{1.15}
\plotone{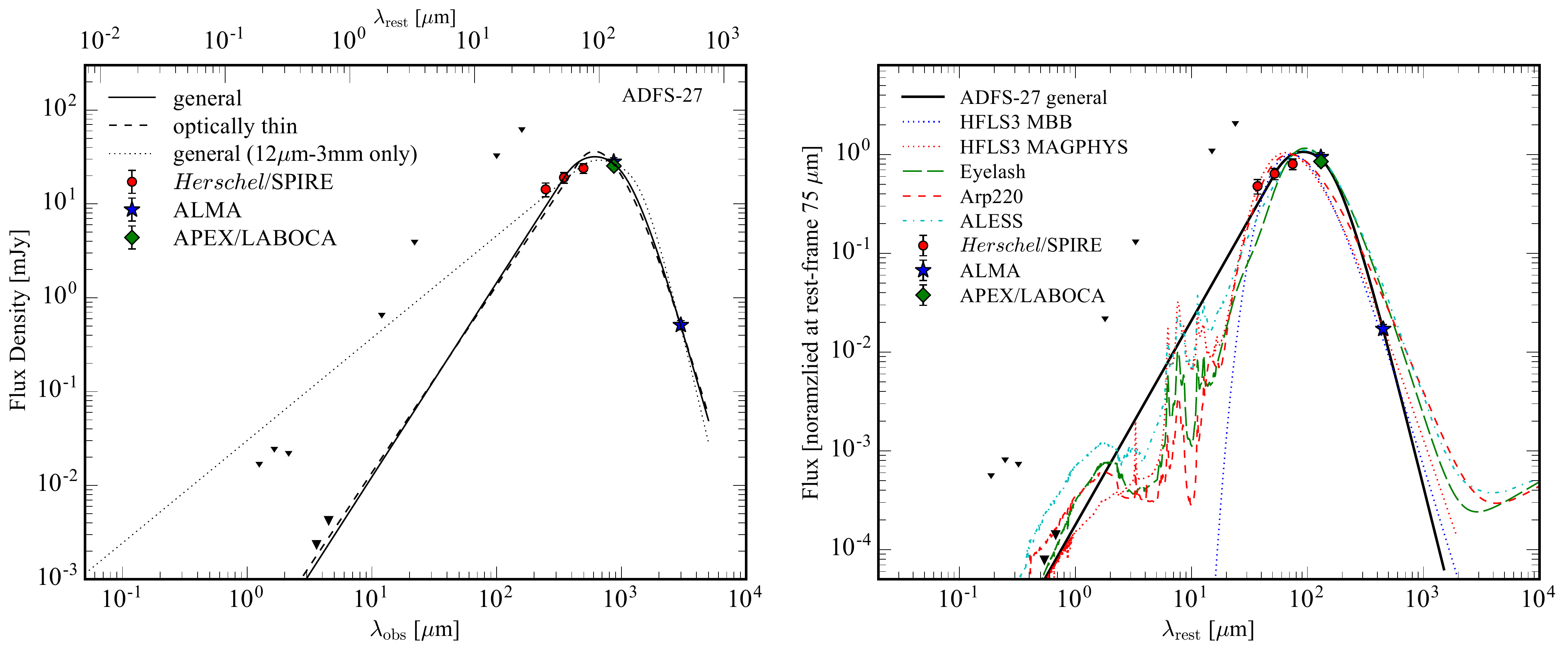}
\vspace{-2mm}

\caption{Spectral energy distribution of \adfs27. The {\em
    Herschel}/SPIRE and APEX/LABOCA photometry were used for the
  initial selection of the source. Upper limits are from {\em
    Spitzer}/IRAC (de-blended fluxes), {\em WISE,} and {\em
    Herschel}/PACS. {\em Left:} The solid line shows our best fit to
  the data. The dashed line shows the best fit assuming optically thin
  emission. The dotted line shows a fit to data at $\geq$12\,$\mu$m
  only. {\em Right:} Same, overplotted with template SEDs for the
  starbursts HFLS3 ($z$=6.34; dotted lines), the Eyelash ($z$=2.33;
  long dashed), and Arp\,220 (dashed), and a composite for ALESS
  sources (dash-dotted; \citealt{riechers13b,cooray14,ivison16}) when
  normalized to the same rest-frame 75\,$\mu$m flux
  density. \label{f4}}
%
\end{figure*}

We detect strong continuum emission at 3\,mm and 870\,$\mu$m at peak
significances of $\sim$39 and 28$\sigma$ toward \adfs27, yielding flux
densities of (0.512$\pm$0.023) and (28.1$\pm$0.9)\,mJy, respectively
(Figs.~\ref{f2}, {\em bottom right} and \ref{f3}, respectively). The
emission is marginally resolved at 3\,mm, and it breaks up into two
components of similar strength separated by 1.49$''$ in the
high-resolution 870\,$\mu$m data, with flux densities of
(15.70$\pm$0.76) and (12.43$\pm$0.56)\,mJy for the northern and
southern components (hereafter:\ \adfsn, or ``mal'' 말, the horse, and
\adfss, or ``yong'' 용, the dragon), respectively.\footnote{Extracted
  from a map tapered to 0.8$''$ resolution.} The two components thus
contain the full single-dish 870\,$\mu$m flux. Both components are
spatially resolved. Two-dimensional Gaussian fitting yields
deconvolved sizes of (0.303$\pm$0.030)$\times$(0.213$\pm$0.027) and
(0.341$\pm$0.031)$\times$(0.146$\pm$0.025)\,arcsec$^2$ for \adfsn\ and
S, respectively. After removal of a bright foreground star, some faint
residual emission is seen at 3.6 and 4.5\,$\mu$m
near the position of \adfs27\ and consistent with the expected flux
levels (Fig.~\ref{f4}), but higher resolution observations would be
required to confirm its mid-infrared detection (Fig.~\ref{f3};
Tab.~\ref{t1}). Given the lack of a candidate lensing galaxy at short
wavelengths or
arc-like structure in the high-resolution ALMA data, there presently
is no evidence for strong gravitational lensing (i.e., flux
magnification factors $\mu_{\rm L}$$\geq$2),
but detailed imaging with the {\em Hubble Space Telescope} would be
required to further investigate the possibility of strong or weak
lensing.


\begin{deluxetable}{ l c l }
\tabletypesize{\scriptsize}
\tablecaption{\adfs27 continuum photometry \label{t1}}
\tablehead{
Wavelength & Flux density\tablenotemark{a} & Telescope \\
 ($\mu$m) & (mJy) & }
\startdata
1.25                   & $<$0.015          & VISTA/VHS \\
1.65                   & $<$0.022          & VISTA/VHS \\
2.15                   & $<$0.020          & VISTA/VHS \\
3.6\tablenotemark{b}   & (2.33$\pm$0.74)$\times$10$^{-3}$ & {\em Spitzer}/IRAC \\
4.5\tablenotemark{b}   & (4.20$\pm$0.82)$\times$10$^{-3}$ & {\em Spitzer}/IRAC \\
12                     & $<$0.6          & {\em WISE} \\
22                     & $<$3.6          & {\em WISE} \\
110                    & $<$30           & {\em Herschel}/PACS \\
160                    & $<$57           & {\em Herschel}/PACS \\
250\tablenotemark{c,d} & 14.3$\pm$2.3    & {\em Herschel}/SPIRE \\
350\tablenotemark{c,d} & 19.1$\pm$2.3    & {\em Herschel}/SPIRE \\
500\tablenotemark{c,d} & 24.0$\pm$2.7    & {\em Herschel}/SPIRE \\
870\tablenotemark{c}   & 25.4$\pm$1.8    & APEX/LABOCA \\
870                    & 28.1$\pm$0.9    & ALMA \\
3000                   & 0.512$\pm$0.023 & ALMA (scan) \\
\enddata 
\tablenotetext{\rm a}{Limits are 3$\sigma$.}
\tablenotetext{\rm b}{Possibly contaminated by foreground sources, and hence, considered as upper limits only in the SED fitting.}
\tablenotetext{\rm c}{Used for initial color/photometric redshift selection.}
\tablenotetext{\rm d}{Uncertainties do not account for confusion noise, which
  is 5.9, 6.3, and 6.8\,mJy (1$\sigma$) at 250, 350, and 500\,$\mu$m,
  respectively \citep{nguyen10}.}
\end{deluxetable}


\subsection{Line Emission}

\begin{figure}
\epsscale{1.15}
\plotone{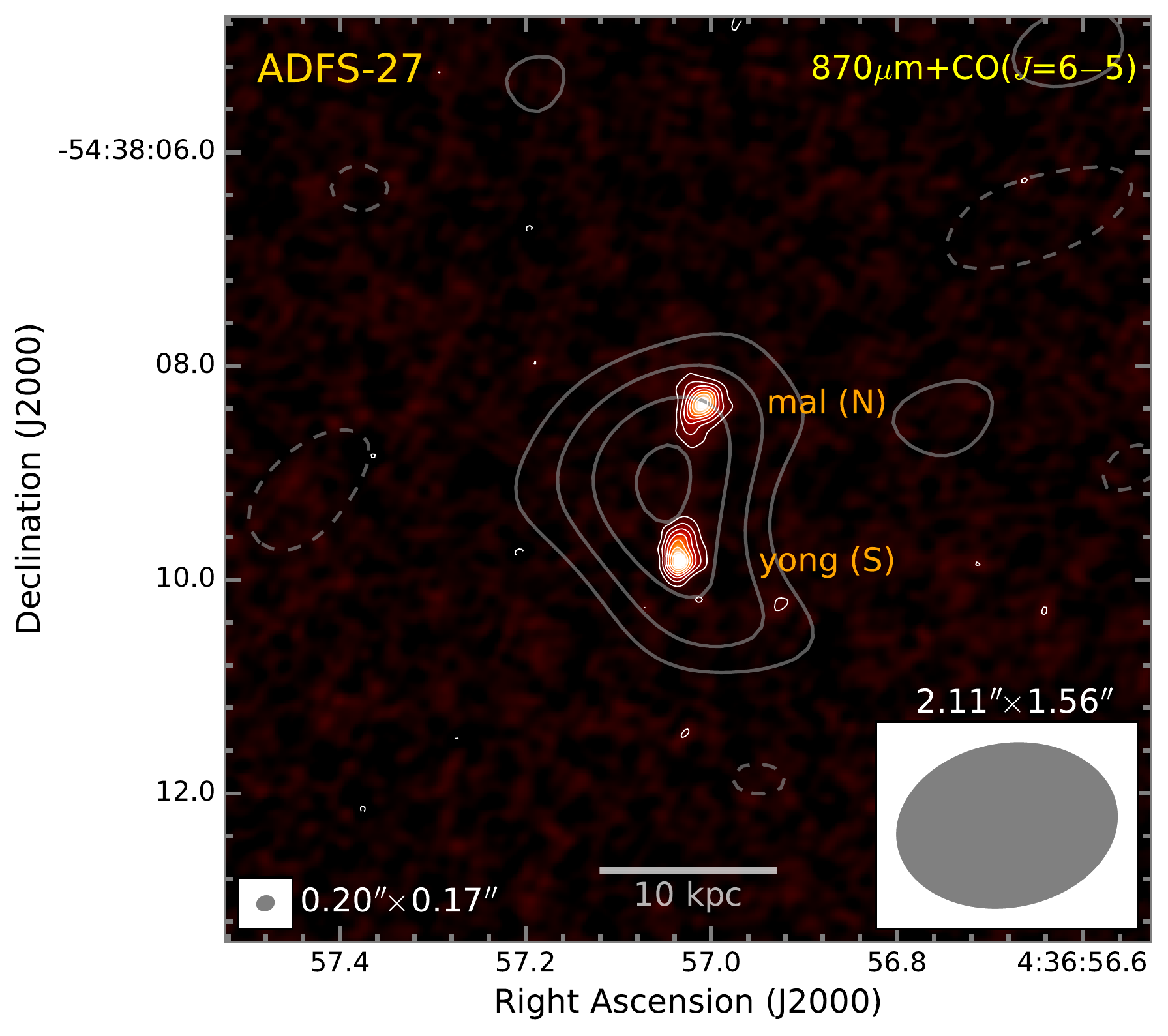}
\vspace{-2mm}

\caption{Overlay of ALMA \fco\ emission (uniform weighting; gray
  contours) on 870\,$\mu$m continuum emission (natural weighting;
  color scale and white contours) for \adfs27. CO velocity range is
  the same as in Fig.~\ref{f2}. CO contours are shown in steps of
  1$\sigma$=0.23\,Jy\,\kms\,beam$^{-1}$, starting at
  $\pm$2$\sigma$. CO beam size is indicated in the bottom right
  corner.  870\,$\mu$m beam size (bottom left corner) and contour
  levels are the same as in Fig.~\ref{f3}. \label{f5}}
%
\end{figure}

A search of the 3\,mm spectral sweep reveals two strong features near
86.6 and 103.9\,GHz detected at $\sim$19 and 12$\sigma$ significance,
respectively. Together with a third, tentative feature near 113.0\,GHz
recovered at 2.3$\sigma$ significance, we obtain a unique (median)
redshift solution at $z$=5.6550$\pm$0.0001, identifying the features
as \eco, \fco, and \water\ emission (Fig.~\ref{f2}, {\em
  top}).\footnote{No spectral lines are detected in the ALMA
  870\,$\mu$m data.} The \water\ line recovery is marginal at best and
near the edge of the spectral range. Thus, an independent confirmation
of this feature is required. The line emission is marginally resolved
on the longest baselines and elongated along the axis that separates
the two continuum source components, and thus, is consistent with
emerging from both sources (Fig.~\ref{f5}). From Gaussian fitting to
the line profiles, we obtain peak flux densities of $S_{\rm
  line}$=(3.89$\pm$0.28), (3.75$\pm$0.43), and (1.55$\pm$0.37)\,mJy at
FWHM linewidths of d$v$=(651$\pm$59), (710$\pm$103), and
(503$\pm$163)\,\kms, respectively.\footnote{The CO line redshifts
  agree within $<$1$\sigma$, where 1$\sigma$=25 and 43\,\kms\ for the
  CO $J$=5$\to$4 and 6$\to$5 lines, respectively. The fit of the
  \water\ line indicates a blueshift by $-$(237$\pm$64)\,\kms\ with
  respect to the \eco\ line, which we consider to be due to limited
  signal-to-noise ratio. Another possible explanation is that the
  H$_2$O emission may preferentially emerge from one of the components
  of \adfs27, assuming a small centroid velocity shift between both
  components. Fixing the line centroid to that of the \eco\ line
  yields $S_{\rm line}$=(0.99$\pm$0.31)\,mJy and
  d$v$=(915$\pm$380)\,\kms, i.e., a $\sim$15\% higher line flux. This
  difference is not significant.}  This implies integrated line fluxes
of (2.68$\pm$0.20), (2.82$\pm$0.34), and
(0.83$\pm$0.22)\,Jy\,\kms\ and line luminosities of $L'_{\rm
  CO}$=(11.96$\pm$0.92) and (8.73$\pm$1.07) and $L'_{\rm
  H_2O}$=(2.17$\pm$0.58)$\times$10$^{10}$\,\lprime, respectively
(Table~\ref{t2}). This yields a \fco/\eco\ line brightness temperature
ratio of $r_{65}$=0.73$\pm$0.10, which is consistent with the average
value for SMGs within the uncertainties ($r_{65}$=0.66;
\citealt{bothwell13}), but significantly lower than that found in the
$z$=5.3 SMG AzTEC-3 ($r_{65}$=1.03$\pm$0.16;
\citealt{riechers10a}). Thus, assuming the average \eco/\aco\ line
brightness temperature ratio for SMGs of $r_{51}$=0.39 \citep{cw13},
we find a \aco\ luminosity of $L'_{\rm
  CO(1-0)}$=3.1$\times$10$^{11}$\,\lprime, i.e., $\sim$50$\times$ that
of Arp\,220 (e.g., \citealt{ds98}).\footnote{Assuming the
  $r_{51}$=0.56 value of AzTEC-3 instead would yield $L'_{\rm
    CO(1-0)}$=2.1$\times$10$^{11}$\,\lprime\ \citep{riechers10a}.} We
also find a \water/\fco\ ratio of $r_{\rm WC}$=0.25$\pm$0.14, which is
$\sim$2.5$\times$ lower than in Arp\,220 and the $z$=6.34 starburst
HFLS3 (\citealt{rangwala11,riechers13b}), and $\sim$1.5$\times$ lower
than in the $z$$\sim$3.5 strongly-lensed starbursts G09v1.97 and
NCv1.143 (\citealt{yang16}; D.~A.\ Riechers et al., in
preparation). This is consistent with a moderate interstellar medium
excitation for a starburst system.

\section{Analysis and Discussion}

\subsection{Spectral Energy Distribution Properties}

To determine the overall spectral energy distribution properties of
\adfs27, we have fit modified black-body (MBB) models to the continuum
data between 1.25\,$\mu$m and 3\,mm
(Fig.~\ref{f4}).\footnote{Confusion noise and flux scale uncertainties
  were added in quadrature where appropriate.}
We adopt the method described by \citet{riechers13b} and
\citet{dowell14}, using an affine-invariant Markov Chain Monte-Carlo
(MCMC) approach, and joining the MBB to a $\nu^\alpha$ power law on
the blue side of the SED peak. We fit optically-thin models, with the
power-law slope $\alpha$, the dust temperature $T_{\rm dust}$, and the
spectral slope of the dust emissivity $\beta_{\rm IR}$ as fitting
parameters, using the observed-frame 500\,$\mu$m flux density as a
normalization factor. We also fit ``general''
models that allow for wavelength-dependent changes in optical depth,
adding the wavelength $\lambda_0$=c/$\nu_0$ where the optical depth
$\tau_\nu$=($\nu$/$\nu_0$)$^{\beta_{\rm IR}}$ reaches unity as an
additional fitting parameter.

The optically-thin fitting procedure yields statistical mean values of
$T_{\rm dust}$=59.9$^{+42.7}_{-33.4}$\,K, $\beta_{\rm
  IR}$=2.3$^{+0.6}_{-1.1}$, and
$\alpha$=6.2$^{+5.0}_{-3.9}$.\footnote{$\alpha$ is only poorly
  constrained by the data.}  The general fit yields mean values of
$\lambda_0$=195$^{+39}_{-41}$\,$\mu$m, $T_{\rm
  dust}$=55.3$^{+7.8}_{-7.6}$\,K, $\beta_{\rm
  IR}$=3.0$^{+0.5}_{-0.5}$, and $\alpha$=9.8$^{+6.7}_{-6.1}$.  The fit
also implies rest-frame infrared (8--1000\,$\mu$m) and far-infrared
(42.5--122.5\,$\mu$m) luminosities of $L_{\rm
  IR}$=2.42$^{+0.48}_{-0.47}$$\times$10$^{13}$\,\lsol\ and $L_{\rm
  FIR}$=1.64$^{+0.27}_{-0.27}$$\times$10$^{13}$\,\lsol,
respectively.\footnote{The measured $L_{\rm IR}$ agrees to within
  $\sim$2\% with independent estimates based on integrating a
  normalized MAGPHYS-based SED template based on the $z$=6.34
  starburst HFLS3 (\citealt{cooray14}), showing that the adopted
  power-law approximation of the short wavelength emission has a minor
  impact on the measured quantities.} Assuming a dust absorption
coefficient of $\kappa_\nu$=2.64\,m$^2$kg$^{-1}$ at 125\,$\mu$m
\citep[e.g.,][]{dunne03}, we also find a dust mass of $M_{\rm
  dust}$=4.4$^{+2.3}_{-2.4}$$\times$10$^9$\,\msol.\footnote{Given the
  limited photometry, the uncertainties may be somewhat
  under-estimated.}  Assuming a \citet{chabrier03} stellar initial
mass function, these parameters suggest a total star formation rate
(SFR) of $\sim$2400\,\msol\,yr$^{-1}$.

Given the limited SED constraints in the rest-frame optical, we obtain
an estimate for the stellar mass $M_{\star}$ of \adfs27\ by
normalizing the MAGPHYS-based SED template of HFLS3 in Fig.~\ref{f4}
to the observed-frame 4.5\,$\mu$m limit. This yields
$M_{\star}$$<$1.2$\times$10$^{11}$\,\msol.

\subsection{Molecular Gas Mass, Gas-to-Dust Ratio, and Gas Depletion Time}

The $L'_{\rm CO(1-0)}$ value of \adfs27\ (based on the adopted
$r_{51}$=0.39) implies a total molecular gas mass of $M_{\rm
  gas}$=2.5$\times$10$^{11}$\,$(\alpha_{\rm
  CO}/0.8)\,(0.39/r_{51})$\,\msol.\footnote{We here adopt a conversion
  factor of $\alpha_{\rm CO}$$=$0.8\,\msol\,(\lprime )$^{-1}$ for
  nearby ultra-luminous infrared galaxies and SMGs
  \citep[e.g.,][]{ds98,tacconi08}.}  Taken at face value, this yields
a gas-to-dust ratio of $M_{\rm gas}$/$M_{\rm dust}$$\simeq$60, which
is comparable to that in the $z$=6.34 starburst HFLS3 and within the
range of values found for nearby infrared-luminous galaxies
\citep{riechers13b,wilson08}, but $\sim$4$\times$ lower than that for
the $z$=5.30 starburst AzTEC-3 \citep{riechers14b}. At its current
SFR, this implies a gas depletion time of $\tau_{\rm dep}$=$M_{\rm
  gas}$/SFR$\simeq$100\,Myr, consistent with the general SMG
population \citep[e.g.,][]{cw13}.


\begin{deluxetable}{ l c c c }

\tabletypesize{\scriptsize}
\tablecaption{Line fluxes and luminosities 
in \adfs27. \label{t2}}
\tablehead{
Transition & $I_{\rm line}$ & $L'_{\rm line}$ & $L_{\rm line}$ \\
           & [Jy\,\kms ] & [10$^{10}$\,K\,\kms\,pc$^2$] & [10$^8$\,\lsol ] }
\startdata
\eco\   & 2.68 $\pm$ 0.20 & 11.96 $\pm$ 0.92 & 7.32 $\pm$ 0.56 \\ 
\fco\   & 2.82 $\pm$ 0.34 &  8.73 $\pm$ 1.07 & 9.24 $\pm$ 1.13 \\ 
\water\tablenotemark{a} & 0.83 $\pm$ 0.22 &  2.17 $\pm$ 0.58 & 2.96 $\pm$ 0.80 
\enddata \tablenotetext{\rm a}{Tentative detection. Independent
  confirmation is required. Quoted uncertainties are from Gaussian
  fitting to the line profile near the edge of the spectral range. We
  consider the true flux uncertainty to be at least $\sim$45\%,
  consistent with line map-based estimates.}
\end{deluxetable}


\vspace{-1cm}

\subsection{Star Formation Rate and Gas Surface Densities, Gas Dynamics, and Conversion Factor}

The apparent 870\,$\mu$m continuum sizes of \adfsn\ (mal) and
\adfss\ (yong) imply physical sizes of
(1.83$\pm$0.18)$\times$(1.28$\pm$0.16) and
(2.05$\pm$0.18)$\times$(0.87$\pm$0.15)\,kpc$^2$ at $z$=5.655, which
are comparable to the $\sim$2.5\,kpc diameters found for other $z$$>$4
dusty starbursts like AzTEC-3, HFLS3, and SGP-38326 at similar
wavelengths \citep{riechers13b,riechers14b,oteo16}. Assuming that
their flux ratios at 870\,$\mu$m are representative at the peak of the
SED, this implies $L_{\rm IR}$ surface densities of $\Sigma_{\rm
  IR}$=7.3 and 7.5$\times$10$^{12}$\,\lsol\,kpc$^{-2}$ and SFR surface
densities of $\Sigma_{\rm SFR}$=730 and
750\,\msol\,yr$^{-1}$\,kpc$^{-2}$, at SFRs of $\sim$1350 and
1070\,\msol\,yr$^{-1}$, respectively, consistent with what is expected
for ``maximum starbursts''
\citep[e.g.,][]{elmegreen99,scoville03,thompson05}. These $\Sigma_{\rm
  SFR}$ values are comparable to those found in other HyLIRGs at
$z$$>$4 like AzTEC-3, HFLS3, and SGP-38326
\citep{riechers13b,riechers14b,oteo16}, but significantly higher than
for the bulk of the DSFG population
\citep[e.g.,][]{tacconi06,bussmann13,bussmann15,hodge16}.

Assuming a common CO linewidth and using the sizes and flux ratio
measured in the 870\,$\mu$m continuum emission, we can obtain
approximate constraints on the dynamical masses $M_{\rm dyn}$ of
\adfsn\ (mal) and \adfss\ (yong) by adopting an isotropic virial
estimator \citep[e.g.,][]{engel10}. We here increase the assumed
source radii by a factor of 1.5 to account for the typical difference
between the measured Gaussian sizes of gas and dust emission in SMGs,
likely caused by decreasing dust optical depth towards the outskirts
of the starbursting regions \citep[e.g.,][]{riechers14b}. We find
$M_{\rm dyn}^{\rm N}$=3.25$\times$10$^{11}$\,\msol\ and $M_{\rm
  dyn}^{\rm S}$=3.66$\times$10$^{11}$\,\msol. Taken at face value, and
conservatively assuming that 100\% of the dynamical mass is due to
molecular gas (i.e., neglecting the potentially major contributions
due to stellar mass and dark matter, and the likely minor
contributions due to dust and black hole masses), this implies an
upper limit of $\alpha_{\rm CO}$$<$2.25\,\msol\,(\lprime )$^{-1}$,
which is consistent with the assumptions made above. This limit drops
to $\alpha_{\rm CO}$$<$1.8\,\msol\,(\lprime )$^{-1}$ when including
the $M_\star$ limit at face value in the estimate. Adopting
$\alpha_{\rm CO}$$=$0.8\,\msol\,(\lprime )$^{-1}$ instead suggests gas
fractions of $f_{\rm gas}$=$M_{\rm gas}$/$M_{\rm dyn}$=0.41 and 0.32
for \adfsn\ and S, respectively. This is comparable to other SMGs
\citep[e.g.,][]{cw13,riechers13b,riechers14b}. Under the same
assumptions, we find gas surface densities of $\Sigma_{\rm gas}^{\rm
  N}$=7.3 and $\Sigma_{\rm gas}^{\rm
  S}$=8.1$\times$10$^{10}$\,\msol\,kpc$^2$. These values are at the
high end of, but consistent with the spatially-resolved
Schmidt-Kennicutt ``star formation law'' \citep[e.g.,][]{hodge15},
providing some of the first constraints on this relation at
$z$$\sim$6.

\section{Conclusions}

We have identified a massive, dust-obscured binary HyLIRG at a
redshift of $z$=5.655, using ALMA. Our target \adfs27\ was selected as
a ``870\,$\mu$m riser'', fulfilling an FIR color criterion of
$S_{250\mu{\rm m}}$$<$$S_{350\mu{\rm m}}$$<$$S_{500\mu{\rm
    m}}$$<$$S_{870\mu{\rm m}}$. Among 25 {\em Herschel}-red sources
(i.e., ``500\,$\mu$m risers'', fulfilling $S_{250\mu{\rm
    m}}$$<$$S_{350\mu{\rm m}}$$<$$S_{500\mu{\rm m}}$)
spectroscopically confirmed to date (e.g., \citealt{riechers13b};
Riechers et al., in prep.) and $\sim$300 photometrically-identified
{\em Herschel}-red sources (\citealt{ivison16,asboth16};
S.~Duivenvoorden et al., in prep.), \adfs27\ is the only point source
to fulfill this additional criterion, implying that such sources are
likely very rare. Of the spectroscopic red sample, all sources are at
$z$$<$5.5 with the exception of HFLS3 at $z$=6.34, which however had
an additional criterion of 1.3$\times$$S_{350\mu{\rm
    m}}$$<$$S_{500\mu{\rm m}}$ applied in its selection
\citep{riechers13b}. \adfs27\ is significantly redder than HFLS3 in
its 870\,$\mu$m/500$\mu$m color (1.06 vs.\ 0.70). Of the 39
spectroscopically confirmed, 1.4+2.0\,mm-selected sample from the SPT
survey, only SPT\,0243--49 at $z$=5.6991 fulfills the
``870\,$\mu$m-riser'' criterion \citep{strandet16}. While not
providing a complete selection of $z$$\gg$5 DSFGs, this shows the
potentially very high median redshifts of such sources, which likely
significantly exceeds that of the parent sample of red
sources.\footnote{SPT\,0459--59 at $z$=4.7993 does not fulfill the
  selection criterion with the revised LABOCA 870\,$\mu$m flux of
  (61$\pm$8)\,mJy found by \citet{spilker16}. However, our discussion
  would remain largely unchanged if we included this source.}  The
apparent submillimeter fluxes of this source are $\sim$3$\times$
higher than those of \adfs27, but SPT\,0243--49 is strongly
gravitationally lensed and intrinsically less than half as bright as
\adfs27\ (having two components of 6.2 and 5.2\,mJy at 870\,$\mu$m;
\citealt{spilker16}). It thus is not a binary HyLIRG.

The overall properties of the binary HyLIRG \adfs27\ are perhaps most
similar to lower-redshift sources like SGP-38326 at $z$=4.425
\citep{oteo16}. It likely represents a major merger of two already
massive galaxies ($>$3$\times$10$^{11}$\,\msol\ each) at $z$$\sim$6
leading to the formation of an even more massive galaxy, and it
contains several billion solar masses of dust that must have formed at
even earlier epochs. Its existence is consistent with previous
findings of an apparently significantly higher space density of
luminous dusty starbursts back to the first billion years of cosmic
time than previously thought, which may be comparable to the space
density of the most luminous quasars hosting supermassive black holes
at the same epochs (e.g.,
\citealt{riechers13b,asboth16,ivison16}). While the flux limits
achieved by the deepest {\em Herschel} SPIRE surveys are perhaps not
sufficiently sensitive to account for the bulk of dusty galaxies at
$z$$>$5, the population uncovered so far could be of key importance
for understanding the early formation of some of the most massive
quiescent galaxies at $z$$\gtrsim$3 (e.g., \citealt{toft14}). Despite
its extreme properties, \adfs27\ is only barely sufficiently bright
and isolated to allow identification in the deep ADF-S SPIRE data. Of
the $>$1000\,deg$^2$ surveyed with SPIRE (e.g., \citealt{oliver12}),
only $\sim$110\,deg$^2$ are sufficiently deep and high quality to
identify ``extremely red'' sources as bright as \adfs27\ without the
aid of strong gravitational lensing. Our results indicate that such
sources are rare, with space densities as low as
9$\times$10$^{-3}$\,deg$^{-2}$ if our measurement is representative,
but they could remain hidden in larger numbers among strongly-lensed
and/or 500\,$\mu$m ``dropout'' samples with strong detections longward
of 850\,$\mu$m, identified in large-area surveys with JCMT/SCUBA-2,
APEX/LABOCA, ACT and SPT, and future facilities like CCAT-prime.

\acknowledgments

The authors wish to thank the anonymous referee for a helpful and
constructive report.  The National Radio Astronomy Observatory is a
facility of the National Science Foundation operated under cooperative
agreement by Associated Universities, Inc. This paper makes use of the
following ALMA data: ADS/JAO.ALMA\#\,2016.1.00613.S and
ADS/JAO.ALMA\#\,2013.1.00001.S. ALMA is a partnership of ESO
(representing its member states), NSF (USA) and NINS (Japan), together
with NRC (Canada) and NSC and ASIAA (Taiwan), in cooperation with the
Republic of Chile. The Joint ALMA Observatory is operated by ESO, AUI/
NRAO and NAOJ. D.R. acknowledges support from the National Science
Foundation under grant number AST-1614213 to Cornell
University. T.K.D.L. acknowledges support by the NSF through award
SOSPA4-009 from the NRAO. This research makes use of data obtained
with {\em Herschel,} an ESA space observatory with science instruments
provided by European-led Principal Investigator consortia and with
important participation from NASA, through the HerMES project. HerMES
is a Herschel Key Program utilising Guaranteed Time from the SPIRE
instrument team, ESAC scientists and a mission scientist. This work is
based in part on observations made with the {\em Spitzer Space
  Telescope,} which is operated by the Jet Propulsion Laboratory,
California Institute of Technology under a contract with NASA. This
publication made use of data products from the {\em Wide-field
  Infrared Survey Explorer,} which is a joint project of the
University of California, Los Angeles, and the Jet Propulsion
Laboratory/California Institute of Technology, funded by the National
Aeronautics and Space Administration. This work is based on
observations made with APEX under Program ID:\ M-090.F-0025-2012, and
also based on observations obtained as part of the VISTA Hemisphere
Survey, ESO Progam 179.A-2010 (PI:\ McMahon).

\vspace{5mm}
\facilities{ALMA, APEX(LABOCA), Herschel(PACS and SPIRE), Spitzer(IRAC), WISE, ESO:VISTA}

\bibliographystyle{yahapj}
\bibliography{ref.bib}


\end{document}